\documentclass[aps,prl,superscriptaddress,twocolumn]{revtex4-1}
\usepackage{graphicx}
\usepackage{amsmath}
\usepackage{amssymb}
\usepackage{color}
\usepackage{bm}% bold math

\begin{document}
\title{Deconfined Quantum Phase Transition of a Higher-Order Symmetry-Protected Topological State}

\author{Chen Peng}
\affiliation{Department of Physics, Renmin University of China, Beijing 100872, China}
\author{Long Zhang}
\email{longzhang@ucas.ac.cn}
\affiliation{Kavli Institute for Theoretical Sciences and CAS Center for Excellence in Topological Quantum Computation, University of Chinese Academy of Sciences, Beijing, 100190, China}
\author{Zhong-Yi Lu}
%\email{}
\affiliation{Department of Physics, Renmin University of China, Beijing 100872, China}

\date{\today}

\begin{abstract}
A higher-order (HO) symmetry-protected topological (SPT) state can be realized in a plaquette-modulated square lattice antiferromagnet, which hosts a spin-$1/2$ degenerate mode on each corner of the lattice with open boundaries. In this work, we show with the field-theoretic analysis and quantum Monte Carlo simulations that the plaquette modulation can drive a direct topological quantum phase transition from the HOSPT to a trivial disordered phase across the deconfined quantum critical point (DQCP) between the antiferromagnetic (AF) order and the valence bond solid (VBS) order, thus the DQCP is a multicritical point bridging both the AF-VBS transition and the topological transition of the HOSPT phase. Our work thus reveals the ubiquitous duality between topological transitions of SPT phases and DQCPs.
\end{abstract}

\maketitle

\emph{Introduction.}---Continuous quantum phase transitions (QPTs) beyond the Landau paradigm of spontaneous symmetry breaking have greatly deepened our understanding of quantum many-body systems. A distinct class of non-Landau transitions is the QPTs of topological states of matter, among which are those between symmetry-protected topological (SPT) phases without breaking any symmetry \cite{Gu2009, Pollmann2010, Chen2010}. Another class is the deconfined quantum critical points (DQCPs) between ordered phases that break incompatible symmetries, the prototype of which is the QPT from the antiferromagnetic (AF) order breaking the spin rotation symmetry to the valence bond solid (VBS) phase breaking the lattice symmetries in frustrated antiferromagnets \cite{Senthil2004a, Senthil2004b}.

Certain examples in both classes of non-Landau transitions can be captured by the nonlinear $\sigma$ models (NLSMs) with topological terms, thus are dual to each other in the low-energy limit \cite{He2016b, You2016, Wang2017a, Qin2017}. Here is a brief argument. First, an SPT phase can be described by the NLSM in the strong-coupling phase with a topological $\theta$-term, whose phase angle is an integer multiple of $2\pi$ \cite{Haldane1985, Affleck1985a, Xu2013e, Bi2015}. Its transition to a trivial phase may be achieved by tuning the phase angle, which may lead to a quantum critical point (QCP) \cite{Affleck1985a, Martin-Delgado1996, Xu2013c}. On the other hand, the topological term also captures the ``topological intertwinement'' of different order parameters; that is, the topological defect of one symmetry-breaking order carries the quantum number of another order, and vice versa, thus restoring one symmetry by proliferating its topological defects spontaneously breaks the other symmetry and induces a DQCP in between \cite{Levin2004, Tanaka2005, Senthil2006}. The duality of an SPT transition \cite{He2016b} and the easy-plane AF-VBS DQCP was proposed theoretically \cite{Xu2015, You2016, Wang2017a}, and nontrivial scaling relations of their critical exponents derived from the duality were confirmed numerically \cite{Qin2017}.

\begin{figure}[b]
\centering
\includegraphics[width=0.48\textwidth]{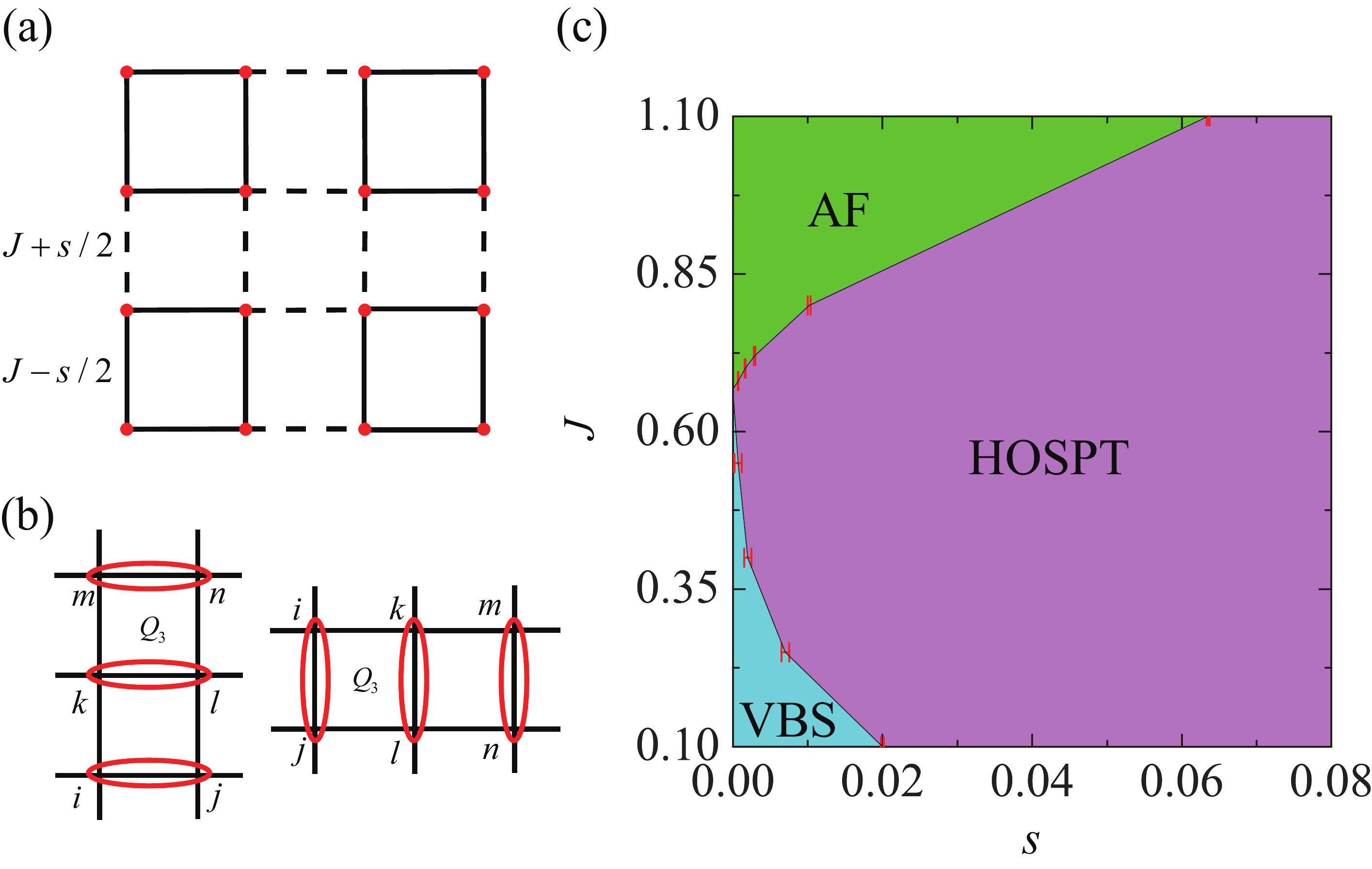}
\caption{(a) The plaquette-modulated square lattice. The solid and dashed lines stand for intra- and inter-UC bonds with Heisenberg interaction strengths $J-s/2$ and $J+s/2$, respectively. (b) Patterns of the multispin interactions in the $Q_{3}$ term. The projection operators in Eq. (\ref{eq:q3}) are marked by red circles. (c) Phase diagram of the plaquette-modulated $J$-$Q_{3}$ model obtained from quantum Monte Carlo simulations. $Q_{3}$ is set to be unity.}
\label{fig:lat}
\end{figure}

In this work, we study the QPTs of a higher-order (HO) SPT phase in the plaquette-modulated square lattice [Fig. \ref{fig:lat} (a)], in which each unit cell (UC) contains four spin-$1/2$ sites. The inter- and intra-UC AF Heisenberg interaction strengths are $J+s/2$ and $J-s/2$, respectively,
\begin{equation}
H_{J}=(J+s/2)\sum_{\langle ij\rangle:\mathrm{inter}}\vec{S}_{i}\cdot\vec{S}_{j} + (J-s/2)\sum_{\langle ij\rangle:\mathrm{intra}} \vec{S}_{i}\cdot\vec{S}_{j}.
\end{equation}
For $s\simeq 2J>0$, the ground state is a direct product of plaquette singlets of the four-spin clusters joined by inter-UC bonds. With an open boundary condition, spins on the edges are dimerized into singlets as well, but there is a dangling spin-$1/2$ mode at each corner, whose two-fold degeneracy is protected by the spin rotation symmetry. The presence of the dangling corner modes is protected either by the lattice $C_{4}$ rotation or the diagonal reflection $R_{xy}$ symmetry, and is the signature of the HOSPT phase \cite{Song2017a, You2018a, Dubinkin2019}. The HOSPT phase persists in an extended range of the plaquette modulation $s$ down to a QCP $s_{c}>0$. For $-s_{c}<s<s_{c}$, the bulk spin gap closes and the ground state has a long-range AF order. For $s<-s_{c}$, the ground state is composed of intra-UC plaquette singlets, but there are not any corner modes, thus it forms a trivial disordered phase.

Is it possible to realize a continuous QPT from the HOSPT to the trivial disordered phase? We consider the following multispin interaction, called the $Q_{3}$ term \cite{Sandvik2007, Lou2009a},
\begin{equation}
H_{Q}=-Q_{3}\sum_{[ijklmn]}P_{ij}P_{kl}P_{mn},
\label{eq:q3}
\end{equation}
in which $P_{ij}=1/4-\vec{S}_{i}\cdot\vec{S}_{j}$ is the projection operator into the spin singlet state. The summation is taken over all triples of parallel bonds in the square lattice shown in Fig. \ref{fig:lat} (b). The $Q_{3}$ term favors the columnar VBS order and drives an AF-VBS DQCP in the absence of the plaquette modulation \cite{Sandvik2007, Lou2009a}.

In this work, we show that the plaquette modulation ($s\neq 0$) drives the DQCP directly into the HOSPT or the trivial disordered phase depending on the sign of $s$. Therefore, the DQCP is a quantum multicritical point bridging both the AF-VBS transition and the topological transition from the HOSPT to the trivial disordered phase. The phase diagram of the plaquette-modulated $J$-$Q_{3}$ model obtained from numerical simulations is shown in Fig. \ref{fig:lat} (c), which is fully consistent with our field-theoretic analysis. The critical exponents of the topological transition are connected to those of the AF-VBS DQCP by nontrivial scaling relations. Our work thus reveals the ubiquitous duality between topological transitions of SPT phases and the DQCPs.

\emph{Field-theoretic analysis.}---The AF-VBS DQCP is captured by the SO(5) NLSM action with a topological Wess-Zumino-Witten (WZW) term \cite{Tanaka2005, Senthil2006},
\begin{equation}
S_{0}=\frac{1}{2g}\int d^{2}xd\tau (\partial_{\mu}\phi)^{2} -2\pi i\Gamma[\phi]+\ldots
\label{eq:so5}
\end{equation}
The superspin field $\phi=(\vec{n},v_{-},v_{+})\in S^{4}$ is composed of the collinear AF order $\vec{n}(\vec{r})\propto (-1)^{r_{x}+r_{y}}\vec{S}(\vec{r})$, and the VBS order $v_{\pm}=v_{x}\pm v_{y}$, in which $v_{\alpha}(\vec{r})\propto (-1)^{r_{\alpha}}(\vec{S}(\vec{r})\cdot \vec{S}(\vec{r}+\hat{\alpha})-\vec{S}(\vec{r}+\hat{\alpha})\cdot \vec{S}(\vec{r}+2\hat{\alpha}))$ ($\alpha=x,y$) is the columnar VBS order. Compared with the original form introduced in Refs. \cite{Tanaka2005, Senthil2006}, the VBS order is rotated into the $v_{\pm}$ basis for later convenience. Under the spin rotation $\mathcal{R}$, the plaquette-centered $C_{4}$ rotation and the diagonal reflection $R_{xy}$, the order parameters transform as
\begin{align}
\mathcal{R}:~& \vec{n}\mapsto R\vec{n},\quad v_{\pm}\mapsto v_{\pm}, \label{eq:transR}\\
C_{4}:		~& \vec{n}\mapsto -\vec{n},\quad v_{\pm}\mapsto \pm v_{\pm}, \label{eq:transC4}\\
R_{xy}:		~& \vec{n}\mapsto \vec{n},\quad v_{\pm}\mapsto \pm v_{\pm}, \label{eq:transRxy}
\end{align}
in which $R$ is the vector representation of the spin rotation.

The first term in Eq. (\ref{eq:so5}) is the SO(5)-symmetric NLSM action, and the ellipses denote higher-order terms explicitly breaking the SO(5) symmetry down to the microscopic SO(3)$\times C_{4}$ symmetry, which are irrelevant at the DQCP as shown by the emergent SO(5) symmetry observed in numerical simulations \cite{Nahum2015a}. The WZW term $\Gamma[\phi]$ is given by
\begin{equation}
\Gamma[\phi]= \frac{3}{8\pi^{2}}\int_{0}^{1}du\int d^{2}x d\tau \epsilon_{abcde}\phi_{a}\partial_{u}\phi_{b}\partial_{x}\phi_{c}\partial_{y}\phi_{d}\partial_{\tau}\phi_{e},
\end{equation}
in which the superspin field is lifted to a continuous field defined on the spacetime extended by an auxiliary parameter $u\in [0,1]$. The lifting is an arbitrary continuous mapping satisfying that
\begin{equation}
\phi(\vec{r},\tau,u=0)=(\vec{0},0,1),\quad \phi(\vec{r},\tau,u=1)=\phi(\vec{r},\tau).
\end{equation}

The plaquette modulation in the Heisenberg interaction couples to the plaquette VBS order $v_{+}$, i.e., the last component of the superspin field,
\begin{equation}
S_{\mathrm{PM}}=s\int d^{2}x d\tau v_{+},
\end{equation}
and polarizes the mean-field ground state in the $v_{+}$ direction, $(\vec{0},0,-\mathrm{sgn} s)$, and gaps out the transverse fluctuations. Taking the following particular choice of the lifting of the superspin field, $\phi=(\sin (u\theta)\hat{\Omega},\cos(u\theta))$, in which $\theta$ is the azimuthal angle of $S^{4}$, and $\hat{\Omega}\in S^{3}$ denotes the normalized transverse fluctuations around the mean-field ground state, we shall derive an effective action of $\hat{\Omega}$. Noting that at the ground state, $\theta$ approaches either $0$ ($s<0$) or $\pi$ ($s>0$) in the whole spacetime, and integrating out $u$, we find that the WZW term reduces into
\begin{equation}
\Gamma[\phi]=\frac{1}{4}(2-3\cos \theta+\cos ^{3}\theta)\Theta[\hat{\Omega}]=
\begin{cases}
0,\quad s<0,\\
\Theta[\hat{\Omega}],\quad s>0,
\end{cases}
\end{equation}
in which $\Theta[\hat{\Omega}]$ is the O(4) $\theta$-term,
\begin{equation}
\Theta[\hat{\Omega}]=\frac{1}{2\pi^{2}}\int d^{2}x d\tau \epsilon_{abcd}\Omega_{a}\partial_{x}\Omega_{b}\partial_{y}\Omega_{c}\partial_{\tau}\Omega_{d}.
\end{equation}
Therefore, the fluctuations of $\hat{\Omega}$ are captured by the O(4) NLSM in the gapped phase for $s<0$, which corresponds to a trivial disordered phase; while for $s>0$, there is an extra $O(4)$ $\theta$-term with a $2\pi$ phase angle. As shown in Ref. \cite{You2018a}, this $\theta$-term together with the $C_{4}$ and the $R_{xy}$ transformations on $\hat{\Omega}$ given in Eqs. (\ref{eq:transC4}) and (\ref{eq:transRxy}), guarantee a degenerate spin-$1/2$ mode at each corner of the lattice with open boundaries, thus capture the essential feature of the HOSPT phase. This indicates that the DQCP also realizes a direct topological transition between the HOSPT and the trivial disordered phase.

The topological transition is driven by $\phi_{5}=v_{+}$, thus its correlation length exponent $\nu_{s}$ is related to the scaling dimension $\Delta_{s}$ of the superspin field $\phi$ at the DQCP. This implies the scaling relation between $\nu_{s}$ and the anomalous dimension $\eta$ of the AF and the VBS order at the DQCP,
\begin{equation}
\Delta_{s}=d-1/\nu_{s}=(d-2+\eta)/2,
\end{equation}
in which $d=3$ is the spacetime dimension assuming that the dynamical exponent $z=1$.

In the absence of the plaquette modulation, the AF order $\langle \vec{n}\rangle\neq 0$ in the large $J/Q_{3}$ regime breaks the spin rotation symmetry, while the columnar VBS order $\langle v_{-}\rangle\neq 0$ in the small $J/Q_{3}$ regime \cite{Lou2009a} breaks the $C_{4}$ rotation symmetry down to $C_{2}$. These symmetry-breaking orders are expected to persist in a finite range of the plaquette modulation except precisely at the DQCP. Both symmetries are restored in the HOSPT and the trivial disordered phase. The transitions from the AF and the VBS orders into the disordered phases are expected to belong to the conventional 3D O(3) and the 3D Ising universality class, respectively. The global phase diagram from the field-theoretic analysis agrees with our numerical results shown in Fig. \ref{fig:lat} (c), which will be described in detail below.

\emph{Numerical method.}---The stochastic series expansion (SSE) quantum Monte Carlo (QMC) with the loop update algorithm \cite{Sandvik1992a, Sandvik1999} is adopted to simulate the plaquette-modulated $J$-$Q_{3}$ model on square lattices of side lengths up to $L=112$ with $N=L^{2}$ sites in the periodic boundary condition. $Q_{3}$ is set to be unity in the simulations. The inverse temperature is proportional to the lattice size, $\beta=L$. At least $10^{5}$ Monte Carlo steps are carried out for each parameter and lattice size. We only need to simulate the $s>0$ case, because the $s<0$ case can be obtained from the former by a lattice translation in the diagonal direction $T_{\hat{x}+\hat{y}}$, under which
\begin{equation}
T_{\hat{x}+\hat{y}}:~\vec{n}\mapsto \vec{n},\quad v_{\pm}\mapsto -v_{\pm}.
\end{equation}
Therefore, the phase diagram is symmetric under the reflection $s\leftrightarrow -s$ except that the HOSPT and the trivial disordered phases are interchanged.

The following physical quantities are calculated to map out the phase diagram and to characterize the QPTs. The AF order parameter is defined by
\begin{equation}
m_{z}=\frac{1}{N}\sum_{\vec{r}}(-1)^{r_{x}+r_{y}}S_{z}(\vec{r}).
\end{equation}
The VBS order is defined by $v_{\pm}=v_{x}\pm v_{y}$, and
\begin{equation}
v_{\alpha}=\frac{1}{N}\sum_{\vec{r}}(-1)^{r_{\alpha}}S_{z}(\vec{r})S_{z}(\vec{r}+\hat{\alpha}),\quad\alpha=x,y.
\end{equation}

\begin{figure}[bt]
\centering
\includegraphics[width=0.48\textwidth]{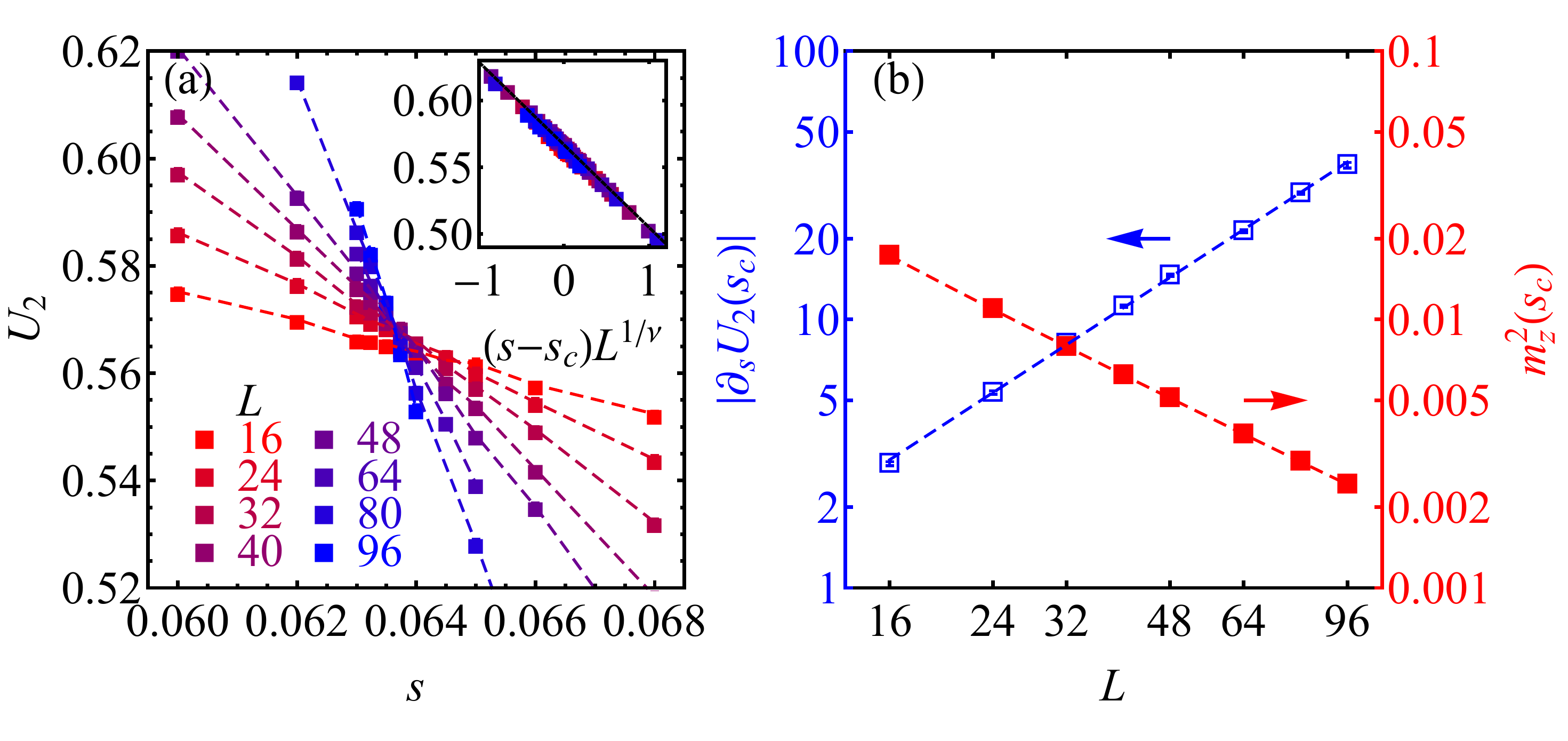}
\caption{(a) The AF Binder cumulant $U_2$ near the AF-HOSPT transition for $J=1.1$. Inset: data collapse in the scaling form, $U_{2}(s,L)=\tilde{u}_{2}((s-s_{c})L^{1/\nu})$ gives an estimate of the QCP, $s_{c}=0.06370(2)$. (b) Finite-size scaling of the slope of $U_2$ and the AF static structure factor $m_{z}^{2}$ at the QCP gives $\nu = 0.701(6)$ and $2\beta/\nu = 1.035(2)$.}
\label{fig:AFM-HOSPT}
\end{figure}

\emph{AF-HOSPT transition.}---We first focus on the AF-HOSPT transition in the $J>J_{c}$ regime, in which $J_{c}=0.667$ is the DQCP in the absence of the plaquette modulation \cite{Lou2009a}. The Binder cumulant of the AF order,
\begin{equation}
U_{2}=\frac{5}{2}\Big(1-\frac{3}{5}\frac{\langle m_{z}^{4}\rangle}{\langle m_{z}^{2}\rangle^{2}}\Big)
\end{equation}
has the following scaling form near a continuous QPT, $U_{2}(s,L)=\tilde{u}_{2}((s-s_{c})L^{1/\nu})$. The data collapse in the inset of Fig. \ref{fig:AFM-HOSPT} (a) indicates a continuous transition, from which the QCP $s_{c}$ is obtained. Moreover, the correlation length exponent $\nu$ can be obtained from the finite-size scaling (FSS) of the slope of $U_{2}$ at the QCP, $\partial_{s}U_{2}(s_{c},L)\propto L^{1/\nu}$. The FSS of the AF static structure factor at the QCP yields another critical exponent $\beta$, $m_{z}^{2}(s_{c},L)\propto L^{-2\beta/\nu}$. Numerical results at $J=1.1$ [Fig. \ref{fig:AFM-HOSPT} (b)] yield $\nu=0.701(6)$ and $2\beta/\nu=1.035(2)$, which are consistent with the 3D O(3) universality class as expected above from the spin rotation symmetry breaking.

The AF-HOSPT phase boundary is obtained from simulations at different $J$ [Fig. \ref{fig:lat} (c)]. The QCP $s_{c}$ is found to shrink to zero as the DQCP is approached from above, $J-J_{c}\rightarrow 0^{+}$. This indicates that an infinitesimal plaquette modulation drives the DQCP into the HOSPT phase in agreement with our field-theoretic analysis.

\begin{figure}[bt]
\centering
\includegraphics[width=0.48\textwidth]{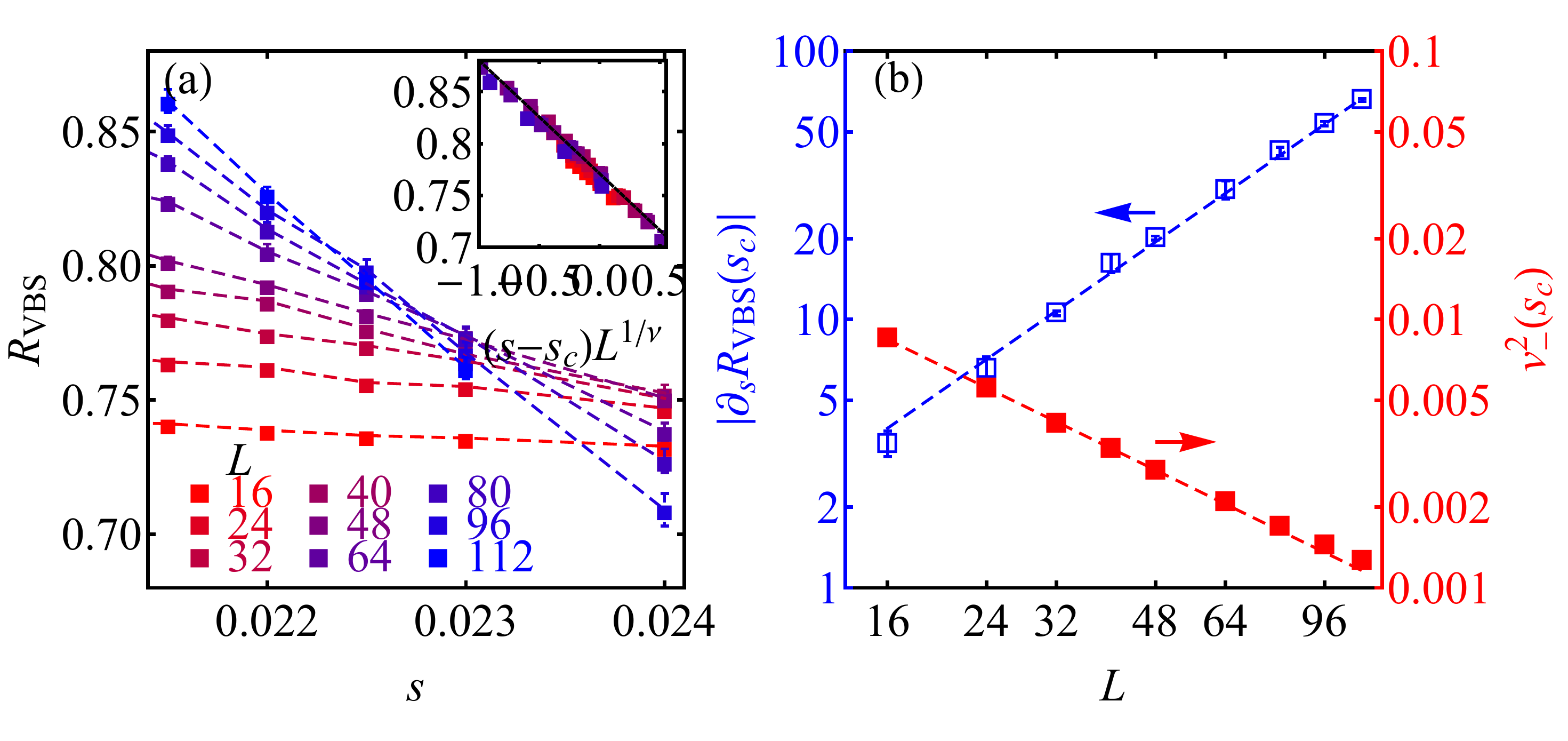}
\caption{(a) The VBS correlation ratio $R_{\mathrm{VBS}}$ close to the VBS-HOSPT transition for $J=0.1$. Inset: data collapse in the scaling form, $R_{\mathrm{VBS}}(s,L)=\tilde{r}_{\mathrm{VBS}}((s-s_{c})L^{1/\nu})$ gives an estimate of the QCP, $s_{c}=0.02296(8)$. (b) Finite-size scaling of the slope of $R_{\mathrm{VBS}}$ and the VBS static structure factor $v_{-}^{2}$ at the QCP gives $\nu = 0.688(9)$ and $2\beta/\nu = 1.019(11)$.}
\label{fig:VBS-HOSPT}
\end{figure}

\emph{VBS-HOSPT transition.}---We turn to the VBS-HOSPT transition in the $J<J_{c}$ regime. The correlation ratio of the columnar VBS order is defined by
\begin{equation}
R_{\mathrm{VBS}}=\frac{\langle v_{-}^{2}\rangle}{\langle |v_{-}(\vec{q})|^{2}\rangle} -1,
\end{equation}
in which $\langle|v_{-}(\vec{q})|^{2}\rangle$ is the static structure factor at $\vec{q}=(2\pi/L,0)$, and $v_{-}(\vec{q})=\frac{1}{N}\sum_{\vec{r}}e^{i\vec{q}\cdot \vec{r}}v_{-}(\vec{r})$. Results at $J=0.1$ are shown in Fig. \ref{fig:VBS-HOSPT}. The data collapse in the inset of Fig. \ref{fig:VBS-HOSPT} (a) also indicates a continuous transition. The FSS of the slope $\partial_{s}R_{\mathrm{VBS}}(s_{c},L)\propto L^{1/\nu}$ and the static structure factor $v_{-}^{2}(s_{c})\propto L^{-2\beta/\nu}$ [Fig. \ref{fig:VBS-HOSPT} (b)] yield $\nu=0.688(9)$ and $2\beta/\nu=1.019(11)$ in agreement with the 3D Ising universality class within error bars, which is expected from the spontaneous breaking of the lattice $C_{4}$ rotation symmetry down to its $C_{2}$ subgroup.

As shown in Fig. \ref{fig:lat} (c), the VBS-HOSPT phase boundary also terminates at the DQCP as $J-J_{c}\rightarrow 0^{-}$. Therefore, the DQCP is established as a quantum multicritical point connecting the AF and the VBS ordered phases as well as the HOSPT and the trivial disordered phases.

\begin{figure}[bt]
\centering
\includegraphics[width=0.48\textwidth]{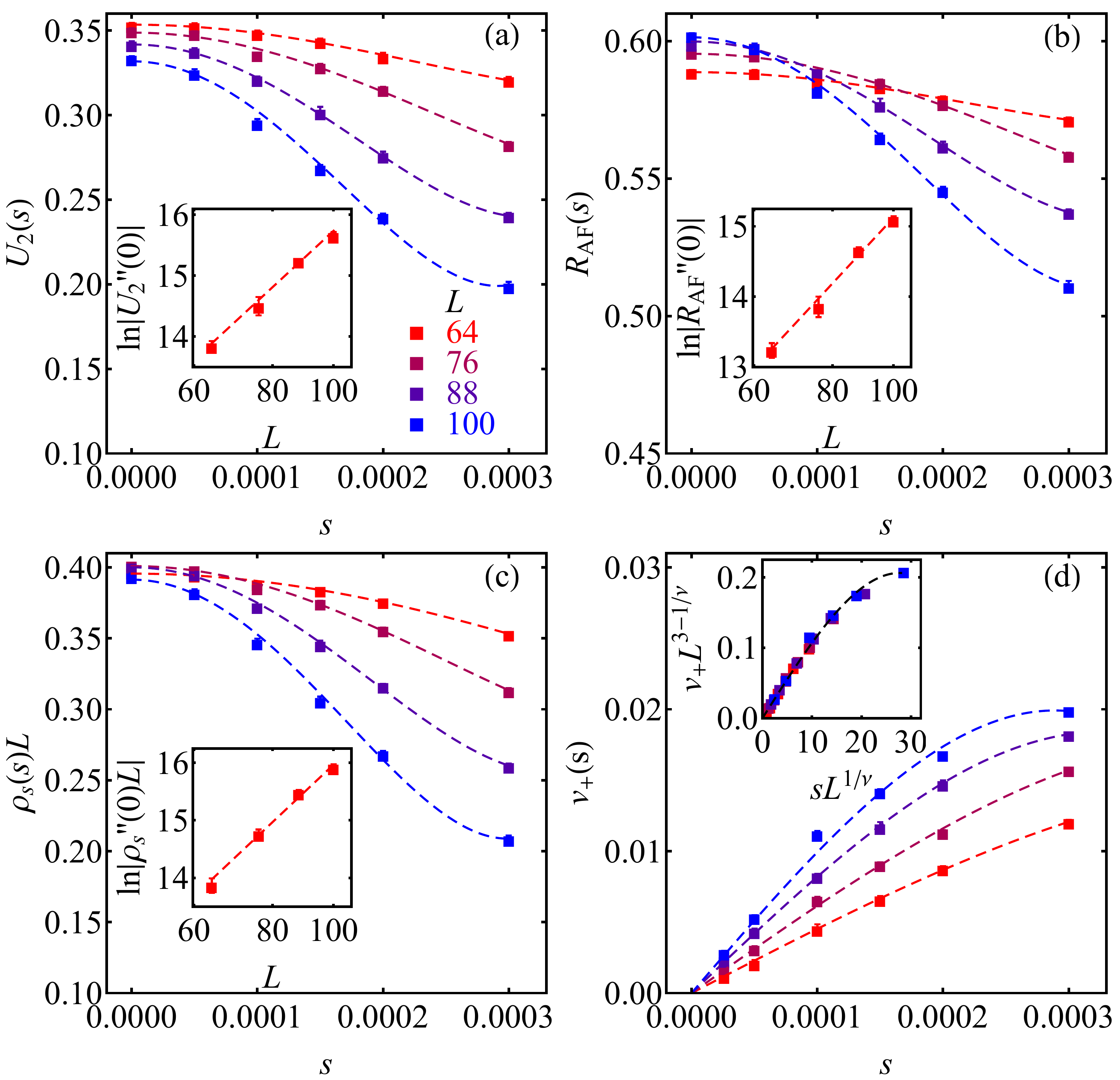}
\caption{Physical quantities close to the topological transition at $J_{c}=0.667$. (a) The AF Binder cumulant $U_2$, (b) the AF correlation ratio $R_{\mathrm{AF}}$, and (c) $\rho_s L$, in which $\rho_{s}$ is the spin stiffness. Dashed lines are the polynomial function fitting with even powers up to $s^{4}$. The insets show the fitting according to Eq. (\ref{eq:ddq}), which gives $\nu_{s} = 0.48(4)$, $0.48(3)$ and $0.45(3)$, respectively. (d) The plaquette VBS $v_{+}$ induced by the nonzero modulation $s$. The dashed lines are polynomial function fitting including odd powers up to $s^{3}$. The inset shows the data collapse to the scaling form in Eq. (\ref{eq:vfss}), which yields $\nu_{s}=0.402(7)$. This is apparently smaller than the previous estimates by a few standard deviations, which might be attributed to systematic errors underestimated in the data collapse fitting procedure. Our final estimate of the critical exponent is $\nu_{s}=0.45(3)$.}
\label{fig:DQCP}
\end{figure}

\emph{Topological transition across the DQCP.}---Both the AF-HOSPT and the VBS-HOSPT phase boundaries terminate at the DQCP, indicating that the DQCP is a multicritical point bridging both the AF-VBS transition and the topological transition from the HOSPT to the trivial disordered phase. While the former has been extensively studied numerically in the literature \cite{Sandvik2007, Lou2009a, Shao2016a, Sandvik2020}, we shall fix $J=J_{c}$ and focus on the topological transition driven by the plaquette modulation $s$.

In the vicinity of the DQCP, all dimensionless quantities under the scaling transformation, such as the AF Binder cumulant $U_{2}(s,L)$, the AF correlation ratio $R_{\mathrm{AF}}(s,L)$, and $\rho_{s}(s,L)L$, in which $\rho_{s}(s,L)$ is the spin stiffness, satisfy the scaling form $Q(s,L)=\tilde{q}(sL^{1/\nu_{s}})$, even if there is not any long range order on either side of the topological transition. Moreover, $\tilde{q}(x)$ is an even function of its argument because $s$ changes sign under the diagonal translation $T_{\hat{x}+\hat{y}}$ while the above quantities do not. Therefore, $\nu_{s}$ can be obtained from fitting
\begin{equation}
\partial_{s}^{2}Q(s,L)|_{s=0}=L^{2/\nu_{s}}\tilde{q}''(0).
\label{eq:ddq}
\end{equation}
Numerical results for these quantities are shown in Fig. \ref{fig:DQCP} (a)--(c), and yield consistent estimates of $\nu_{s}$, which are listed in the figure caption.

The plaquette VBS order $v_{+}$ induced by a nonzero modulation $s$ has the scaling form, 
\begin{equation}
v_{+}(s,L)=L^{-(d-1/\nu_{s})}\tilde{v}_{+}(sL^{1/\nu_{s}}),
\label{eq:vfss}
\end{equation}
and $\tilde{v}_{+}(x)$ is an odd function. Data collapse in this scaling form is shown in the inset of Fig. \ref{fig:DQCP} (d). The estimate of $\nu_{s}$ roughly agrees with the above results within a few standard deviations. Our final estimate of the correlation-length exponent is $\nu_{s}=0.45(3)$. The numerical conformal bootstrap on the possible continuous DQCP gives a lower bound, $\nu_{s}\ge 0.446$ \cite{Nakayama2016, Poland2019}, and our estimate lies marginally on this bound; however, we cannot draw a definite conclusion due to the relatively large error bar in our result.

\begin{figure}[bt]
\centering
\includegraphics[width=0.48\textwidth]{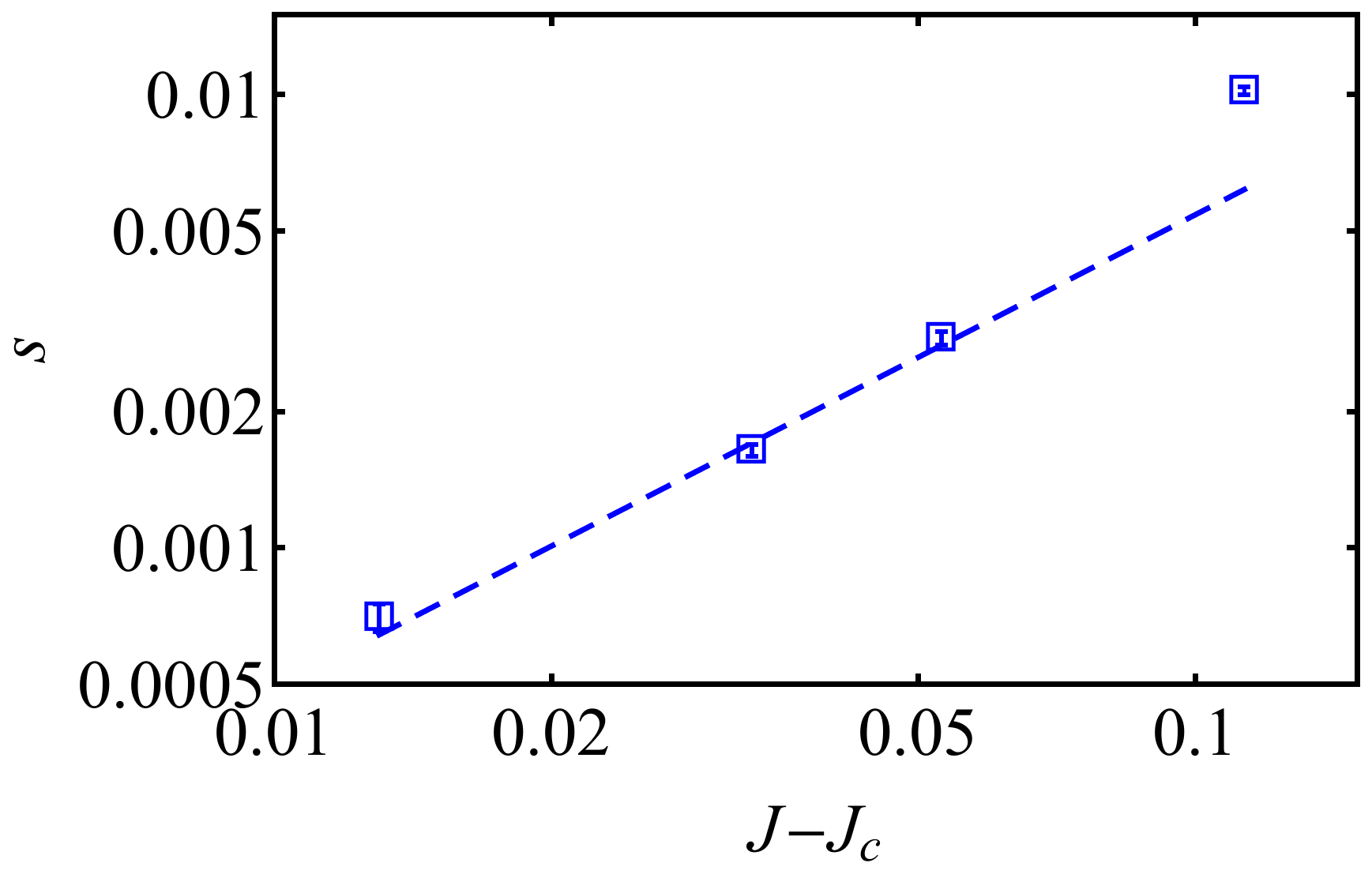}
\caption{The AF-HOSPT phase boundary $J_{c}(s)$ versus the plaquette modulation $s$ in the vicinity of the DQCP. We obtain $\nu_J/\nu_{s}= 1.04(11)$ from fitting the power law in Eq. (\ref{eq:boundary}).}
\label{fig:nu-from-phase-boundary}
\end{figure}

Finally, in the close vicinity of the multicritical point, the AF-HOSPT phase boundary $(s,J_{c}(s))$ should obey the scaling form,
\begin{equation}
s\propto |J_{c}(s)-J_{c}|^{\nu_{J}/\nu_{s}},
\label{eq:boundary}
\end{equation}
in which $\nu_{J}$ is the correlation-length exponent of the AF-VBS transition. This is because in the $(s,J)$ plane, the phase boundary is given by a particular renormalization group (RG) trajectory passing the fixed point of the AF-HOSPT transition, which has the above scaling form near the multicritical point at $(0,J_{c})$. The power-law fitting of the phase boundaries close to the DQCP shown in Fig. \ref{fig:nu-from-phase-boundary} gives $\nu_{J}/\nu_{s}=1.04(11)$, which combined with our estimate of $\nu_{s}$ gives $\nu_{J}=0.47(6)$, in agreement with the recent high-precision numerical estimate, $\nu_{J}=0.455(2)$ \cite{Sandvik2020}.

\emph{Conclusion.}---In summary, we have studied the deconfined topological transition from the HOSPT to the trivial disordered phase in a plaquette-modulated square lattice antiferromagnet with combined efforts of the field-theoretic analysis and numerical simulations. We have shown that an infinitesimal plaquette modulation drives the DQCP into either the HOSPT or the trivial disordered phase depending on the sign of the modulation, thus the DQCP is a multicritical point bridging both the AF-VBS transition and the topological transition of the HOSPT phase. This further reveals the ubiquitous duality between the DQCPs and the topological transitions of SPT phases.

\begin{acknowledgments}
L.Z. is grateful to helpful discussions with Z.-C. Gu and N. Su. C.P. and Z.Y.L. are supported by the National Natural Science Foundation of China (Grants No. 11934020 and No. 11774422). L.Z. is supported by the National Key R\&D Program (No. 2018YFA0305800), the National Natural Science Foundation of China (No. 11804337), CAS Strategic Priority Research Program (No. XDB28000000) and CAS Youth Innovation Promotion Association.
\end{acknowledgments}

\bibliography{library}
\end{document}